\documentclass[aps,prd,eqsecnum,amsmath,nofootinbib,preprintnumbers]{revtex4}%
\usepackage{amsmath,amsthm,amssymb}
\usepackage{amsfonts,amssymb}
\usepackage{graphicx}
\usepackage{xcolor}
\usepackage{float}
\definecolor{wred}{rgb}{0,0.618,0.0}
\definecolor{wblue}{rgb}{.0,0.0,0.618}
\definecolor{wgreen}{rgb}{.0,0.618,0.0}
\usepackage{tikz}
\usepackage{multirow}
\usetikzlibrary{decorations.pathmorphing,calc}
\usepackage{bm}
\usepackage[
colorlinks=true,
filecolor=black,
linkcolor=wblue,
citecolor=wgreen,
pdfstartview=FitH,
pdfpagemode=None,
bookmarksopen=true,
urlcolor=blue
]{hyperref}

\begin{document}
	\preprint{\hfill {\small {1}}}
\title{New Anisotropic Gauss-Bonnet Black Holes in Five Dimensions at the Critical Point}
\author{Yuxuan Peng}
\email{yxpeng@alumni.itp.ac.cn}
\affiliation{Department of Physics, School of Science, East China University of Technology, Nanchang, Jiangxi 330013, P.R. China}
\affiliation{CAS Key Laboratory of Theoretical Physics, Institute of Theoretical Physics, Chinese Academy of Sciences, Beijing 100190, P.R. China}

\begin{abstract}
	We obtain new vacuum static black hole solutions with anisotropic horizons in Einstein-Gauss-Bonnet gravity with a negative cosmological constant in five dimensions.
	The translational invariance along one direction on the $3$-dimensional horizon cross section is broken.
	The Gauss-Bonnet coupling $\alpha$ is at the critical point where there is one single AdS vacuum.
	These solutions does not appear in the form of a warped product, i.e. they lack a common warping factor, and the metric contains $2$ arbitrary functions, $h(r)$ of the radial coordinate $r$ and $H(y)$ of the horizon coordinate $y$ --- some degeneracy in the metric.
	The nontrivial horizon and the degeneracy may be closely related to the critical value of $\alpha$.
	We introduce the process of obtaining the solutions and some of their properties, and also prove a uniqueness theorem for the case when there is a common warping factor for the rest two directions.

\end{abstract}

\maketitle

\section{introduction}

When the spacetime dimension is $4$ and the dominant energy condition (DEC), stationary black hole horizon obeys the topology theorem \cite{Hawking:1971vc}.
Black hole horizons can be complicated while those two conditions are violated, i.e. one can consider higher dimensions \cite{Myers:1986un,Emparan:2001wn,Galloway:2005mf,Emparan:2008eg,Obers:2008pj} or break DEC.
by introducing the negative cosmological constant $\Lambda$.
The asymptotically anti-de Sitter (AdS, the maximally symmetric space with negative curvature) black holes are solutions with negative $\Lambda$ and they can have horionzs of $3$ types: sphere with positive curvature, torus with flat geometry and hyperbolic space with negative curvature \cite{Lemos:1994xp,Lemos:1995cm,Birmingham:1998nr}.
These $3$ types appear in arbitrary dimensions.
If one considers generic black holes that are not asymptotically AdS, with negative $\Lambda$, there are even black hole horizons with arbitrary genus $g>1$ \cite{Aminneborg:1996iz}. These are spacetimes locally equal to pure AdS spacetimes.
Moreover, some static, plane symmetric solutions and cylindrically symmetric solutions of Einstein-Maxwell equations with a negative cosmological constant are investigated in Ref \cite{Cai:1996eg}.

It is interesting to explore the possibility of certain nontrivial horizon types in the presence of a negative $\Lambda$.
There is an important example, Ref. \cite{Cadeau:2000tj}, which obtained nontrivial black holes in $5$ dimensions.
These black holes have $3$-dimensional horizons of different types.
These types all belong to the eight $3$-dimensional ``model-geometries'' classified by Thurston \cite{Th1}.
They all admit homogeneous metrics.
These manifolds should be compact if one uses them to construct black hole horizons, and the topologies and the compactification of these eight manifolds are also described in \cite{Cadeau:2000tj}.
Its main results are Einsteinian solutions of black holes with ``Sol'' and ``Nil'' horizons\footnote{According to \cite{Cadeau:2000tj}, there had already been black hole horizons for the first five geometries except $S^1\times S^2$, while ``$SL_2 R$'' horizons are still unknown.}.

What we have been doing is to explore the effect of higher curvature corrections on the horizon geometry topology.
To be exact, we consider $5$-dimensional Einstein-Gauss-Bonnet (EGB for short) gravity theory, a special case of the Lovelock gravity theory \cite{Lovelock:1971yv}, the general second-order covariant gravity theory in dimensions higher than four, with a negative $\Lambda$. 
The Gauss-Bonnet terms are higher curvature terms in the action that can be regarded as corrections from the heterotic string theory \cite{Gross:1986iv,Zumino:1985dp}. 
There had already been spherical, Euclidean and hyperbolic black hole horizons in this theory, shown by the famous paper \cite{Cai:2001dz} by Rong-Gen Cai.
Ref.\cite{Cvetic:2001bk} investigated the charged solutions and their thermodynamics in such a theory.
What we found are static black holes with a $S^1 \times H^2$ horizon and with a ``Solv''-manifold horizon\cite{Peng:2021xwh}.
The ``Solv''-manifold solution has zero entropy and degeneracy an arbitrary function of the radial coordinate $r$ in the metric.
These may be caused by the fact that for this solution the Gauss-Bonnet coupling constant has a critical value actually forbidding kinetic fluctuations around the AdS backgrounds \cite{Fan:2016zfs}.
Meanwhile, at this critical point the two AdS vacuums merge into one single solution, simplifies the algebra greatly.
Besides, for such critical theories, Ref.\cite{Arenas-Henriquez:2019rph} found that the behavior of the mass parameter in some black hole solutions is dramatically different to non-critical cases.

The ``Solv'' solution inspires us to consider the possibilities of similar nontrivial $3$-dimensional horizons.
This paper's main purpose is to break the translational invariance along one direction\footnote{Breaking this symmetry plays a very important role in the holographic condensed matter models with momentum dissipation \cite{Vegh:2013sk,Davison:2013jba,Blake:2013bqa,Blake:2013owa,Andrade:2013gsa,Cai:2014znn,Davison:2014lua,Xu:2015rfa,Cao:2015cza,Cao:2015cti,Chen:2019zlg}.}:
\begin{eqnarray}
	\text{d}x^2+  G(x) \text{d}\Sigma_2^2 
\end{eqnarray}
with $ \mathrm{d}\Sigma_2^2$ the line-element of a $2$-dimensional space with coordinates $y$ and $z$.
On the other hand, the holographic model with the horizon above will possess a boundary metric
\begin{eqnarray}
	\mathrm{d}\hat{s}^2 = -\mathrm{d}t^2 + \text{d}x^2+  G(x) \text{d}\Sigma_2^2 
\end{eqnarray}
which means a field theory living on a curved spacetime background.
This is reminiscent of the quantum systems in gravitational fields, sometimes dynamic spacetimes, for example gravitational waves and expanding universes \cite{Sabin:2014bua,Biasi:2019eap,An:2019opz}.
Nevertheless, our assumption only involves curvature on the spatial slices, not the time direction, therefore it may have different physical meaning and applications.
To explore such cases is by itself interesting and is important for studying strongly-coupled quantum systems in curved spacetime backgrounds.

In section \ref{secansatz} we describe the method of solving the field equations, and we found a group of novel black hole solutions with such kind of anistotropic horizon and some degeneracies shown in section \ref{secsol}.
We also prove a uniqueness theorem for the case when there is a common warping factor for the rest two directions $y, z$ in section \ref{sectheorem}: if the horizon is to be anisotropic, there must be a common warping factor for all the three directions, i.e. the whole horizon manifold, thus leading to known results in Ref.\cite{Dotti:2007az}. Finally we give the conclusion.

\section{Solve the anisotropic black hole horizon}\label{secansatz}

The action of EGB gravity theory with a negative cosmological constant
\begin{eqnarray}
S &= &\frac{1}{16 \pi }\int \mathrm{d}^{d+1} x \sqrt{-g}\left[ R - 2\Lambda + \alpha (R^2 -4 R_{\mu\nu}R^{\mu\nu} + R_{\mu\nu \rho \sigma}R^{\mu\nu \rho \sigma}) \right]
\,,
\end{eqnarray}
where the Newton constant is set to $1$, the cosmological constant $\Lambda<0$, and the Gauss-Bonnet coupling denoted by $\alpha$.
The equations of motion are
\begin{eqnarray}
&&R_{\mu\nu} - \frac{1}{2}R g_{\mu\nu} + \Lambda g_{\mu\nu}\nonumber\\
&&- \alpha\left(4 R_{\mu\rho} R^\rho\,_\nu - 2R R_{\mu\nu}+ 4 R^{\rho \sigma}R_{\mu\rho\nu\sigma} - 2 R_\mu\,^{\rho\sigma\tau}R_{\nu\rho\sigma\tau} + \frac{1}{2} g_{\mu\nu} (R^2 -4 R_{\rho \sigma}R^{\rho \sigma} + R_{\rho\sigma\tau\pi}R^{\rho\sigma\tau\pi})\right)\nonumber\\
&&=0
\,.\label{GBeom}
\end{eqnarray}

We will consider a static black hole solution with an anisotropic horizon.
The induced metric of the anisotropic horizon ansatz is set to be
\begin{eqnarray}\label{anisohorizon}
\text{d}x^2+  G(x) \text{d}\Sigma_2^2 
\end{eqnarray}
with $ \mathrm{d}\Sigma_2^2$ the line-element of a $2$-dimensional space with coordinates $y$ and $z$.
The ansatz line element of the whole spacetime is
\begin{eqnarray}\label{anisoBHansatz}
\mathrm{d}s^2 = -V{(r)}\mathrm{d}t^2 + \frac{1}{V(r)}\mathrm{d}r^2 +f(r)\mathrm{d}x^2 + g(r)G(x)\mathrm{d}y^2 + h(r)G(x)H(y)
\mathrm{d}z^2\,.
\end{eqnarray}

Here $V(r), f(r), g(r)$, $h(r)$, $G(x)$ and $H(y)$ are functions to be determined.
For the expected black hole solution, any constant-$t,r$ surface, including the horizon, will have the form of Eq.(\ref{anisohorizon}), although the functions $f(r), g(r)$ and $h(r)$ may be different.
That is to say, there is no common warping factor $r^2$ like the Schwarzschild-type black holes.
It is this unique setup that allows the new anisotropic solution to exist. 

For the ansatz (\ref{anisoBHansatz}), the $tt\,, rr\,, xx\,, yy\,, zz$ and $rx\,, ry$ components of the equations (\ref{GBeom}) are nonzero.
Firstly look at the cross terms $rx\,, ry$.
\begin{eqnarray}\label{choose}
\text{The $ry$ equations gives}\quad\frac{H'(y) \Big(h(r) g'(r)-g(r) h'(r)\Big) \Big(f(r)-\alpha  f'(r) V'(r)\Big)}{4 f(r) g(r) h(r) H(y)}=0\,\,.
\end{eqnarray}
There are $2$ choices, $h(r) g'(r)-g(r) h'(r)=0$ or $f(r)-\alpha  f'(r) V'(r)=0$.
We can see below that the second choice gives the new solution, and the first choice give nothing new, which will be described in \ref{sectheorem}.

\textbf{The second choice: We choose $f(r)-\alpha  f'(r) V'(r)=0$, and}
\begin{eqnarray}
\text{the $rx$ equation gives}\quad\frac{G'(x) \Big(f(r) g'(r)-g(r) f'(r)\Big) \Big(f(r) h'(r)-h(r) f'(r)\Big)}{2 f(r) g(r) G(x) h(r) f'(r)}\,.
\end{eqnarray}
One can choose $f(r) g'(r)-g(r) f'(r)=0$. Choosing $f(r) h'(r)-h(r) f'(r)=0$ gives actually the same results.
Here we arrive at a point with
\begin{eqnarray}
f(r)-\alpha  f'(r) V'(r)=0\,,\\
f(r) g'(r)-g(r) f'(r)=0\,.
\end{eqnarray}
This means that 
\begin{eqnarray}\label{rel1}
\frac{f'(r)}{f(r)}= \frac{1}{\alpha  V'(r)}\quad\text{and}\quad
f(r) = a g(r)\, (\text{constant $a$ is unimportant})\,.
\end{eqnarray}
For further convenience, we define 
\begin{eqnarray}
f(r)\equiv e^{\mathcal{F}(r)}\,,g(r)\equiv e^{\mathcal{G}(r)}\,,h(r)\equiv e^{\mathcal{H}(r)}\,.
\end{eqnarray}
Then the relations (\ref{rel1}) are
\begin{eqnarray}\label{rel2}
\mathcal{F}'(r)= \frac{1}{\alpha  V'(r)}\quad,\mathcal{G}(r) = \mathcal{F}(r)+\text{constant}\,.
\end{eqnarray}
Under these conditions, the rest nonzero equations are the $tt\,,rr\,,xx\,,yy\,,zz$ components.
The $tt$ equation is
\tiny
\begin{eqnarray}\label{tteq}
\frac{H''(y)}{H(y)} \left(\frac{e^{-\mathcal{F}(r)} V(r)^2}{2 a \alpha  G(x) V'(r)^2}-\frac{e^{-\mathcal{F}(r)} V(r)^2 V''(r)}{a G(x) V'(r)^2}\right)+\frac{H'(y)^2}{H(y)^2} \left(\frac{e^{-\mathcal{F}(r)} V(r)^2 V''(r)}{2 a G(x) V'(r)^2}-\frac{e^{-\mathcal{F}(r)} V(r)^2}{4 a \alpha  G(x) V'(r)^2}\right)\nonumber\\
+\frac{G''(x)}{G(x)} \left(\frac{e^{-\mathcal{F}(r)} V(r)^2}{2 \alpha  V'(r)^2}+\frac{1}{2} \alpha  e^{-\mathcal{F}(r)} V(r) \mathcal{H}'(r) V'(r)-\frac{e^{-\mathcal{F}(r)} V(r)^2 V''(r)}{V'(r)^2}+\alpha  e^{-\mathcal{F}(r)} V(r)^2 \mathcal{H}''(r)+\frac{1}{2} \alpha  e^{-\mathcal{F}(r)} V(r)^2 \mathcal{H}'(r)^2-\frac{1}{2} e^{-\mathcal{F}(r)} V(r)\right)\nonumber\\
+\frac{G'(x)^2}{G(x)^2} \left(-\frac{1}{4} \alpha  e^{-\mathcal{F}(r)} V(r) \mathcal{H}'(r) V'(r)-\frac{1}{2} \alpha  e^{-\mathcal{F}(r)} V(r)^2 \mathcal{H}''(r)-\frac{1}{4} \alpha  e^{-\mathcal{F}(r)} V(r)^2 \mathcal{H}'(r)^2+\frac{1}{4} e^{-\mathcal{F}(r)} V(r)\right)\nonumber\\
\normalsize
{+(\text{terms including $r$ only})=0\,.}
\end{eqnarray}
\normalsize
For this equation to be solved, the variables $r\,,x$ and $y$ should be separated.
First look at the terms with $x$. There are $2$ terms with $G''/G$ and $G'^2/G^2$.
The coefficients of these $2$ terms should be in proportion to separate the variable $x$, therefore
\tiny
\begin{eqnarray}
\mu \left(\frac{e^{-\mathcal{F}(r)} V(r)^2}{2 \alpha  V'(r)^2}+\frac{1}{2} \alpha  e^{-\mathcal{F}(r)} V(r) \mathcal{H}'(r) V'(r)-\frac{e^{-\mathcal{F}(r)} V(r)^2 V''(r)}{V'(r)^2}+\alpha  e^{-\mathcal{F}(r)} V(r)^2 \mathcal{H}''(r)+\frac{1}{2} \alpha  e^{-\mathcal{F}(r)} V(r)^2 \mathcal{H}'(r)^2-\frac{1}{2} e^{-\mathcal{F}(r)} V(r)\right)
\nonumber\\=
-\frac{1}{4} \alpha  e^{-\mathcal{F}(r)} V(r) \mathcal{H}'(r) V'(r)-\frac{1}{2} \alpha  e^{-\mathcal{F}(r)} V(r)^2 \mathcal{H}''(r)-\frac{1}{4} \alpha  e^{-\mathcal{F}(r)} V(r)^2 \mathcal{H}'(r)^2+\frac{1}{4} e^{-\mathcal{F}(r)} V(r)\,,
\end{eqnarray}
\normalsize
with a constant $\mu$, which can be chosen as $-{1}/{2}$ to simplify the expression to
\begin{eqnarray}
\frac{e^{-\mathcal{F}(r)} V(r) \left(2 \alpha  V''(r)-1\right)}{\alpha  V'(r)}=0\,,
\end{eqnarray}
and this equation admits a simple solution
\begin{eqnarray}\label{lapse}
V(r)=\frac{r^2}{4\alpha}+c_1\,,
\end{eqnarray}
with $c_1$ an integration constant. For a black hole $c_1$ should be negative.
Then (\ref{rel1}) will give
\begin{eqnarray}
f(r)=r^2\quad, \text{and}\quad g(r)= r^2\,\quad\text{up to a }.
\end{eqnarray}
\textbf{The arbitrariness of $H(y)$: after fixing the lapse function $V(r)$, the coefficients of the $H(y)$ terms in (\ref{tteq}) are $0$ automatically.
In fact, the other field equations besides (\ref{tteq}) do not say anything meaningful about $H(y)$, therefore $H(y)$ is arbitrary!}

At this moment we solve $G(x)$, according to the variable separation method.
The terms with $G(x)$ in (\ref{tteq}) are now proportional to
\begin{eqnarray}
\frac{G'(x)^2-2 G(x) G''(x)}{G(x)^2}
\end{eqnarray}
which must be a constant.
The solution is
\begin{eqnarray}
G(x) = \left(b_1 e^{\sqrt{M} x} + b_2 e^{-\sqrt{M} x}\right)^2
\end{eqnarray}
with $b_1\,,b_2$ and $M$ constants.
$G(x)$ can be written as $Sinh$ or $Cosh$ functions.
The rest field equations, the $xx\,, yy\,,$ and $zz$ components force 
\begin{eqnarray}\label{criticalalpha1}
\alpha = -\frac{3}{4\Lambda}\,.
\end{eqnarray}
After setting $\alpha = -{3}/{4\Lambda}$ there are still $2$ nonzero field equations, the $tt$ and $rr$ components:
\begin{eqnarray}\label{resteq}
&&-\frac{(M+c_1) \left(\Lambda  r^2-3 c_1\right) \left(\left(2 \Lambda  r^2-6 c_1\right) \mathcal{H}''(r)+\left(\Lambda  r^2-3 c_1\right) \mathcal{H}'(r)^2-4 \Lambda +2 \Lambda  r \mathcal{H}'(r)\right)}{12 \Lambda  r^2}=0\,,\nonumber\\
&&\frac{3 (M+c_1) \left(r \mathcal{H}'(r)-2\right)}{2 \Lambda  r^4-6 c_1 r^2}=0
\end{eqnarray}
and $c_1=-M$ let them both vanish\footnote{If $c_1\neq-c$ then we must set $r \mathcal{H}'(r)-2$, then $\mathcal{H}'(r)=\ln(r^2)$, therefore $h(r)=r^2= f(r)\,, g(r)$ and this case does not give a new solution as discussed in \ref{sectheorem}.}.

\textbf{The arbitrariness of $h(r)$: after setting $ c_1=-M$ the field equations are satisfied, but according to the equations (\ref{resteq}) the terms including $\mathcal{H}(r)$, i.e. $h(r)$ vanishes. Therefore $h(r)$ is arbitrary!}
In the section below we summarize the solution just found and discuss some of its properties.

\section{new black hole solution with an anisotropic horizon}\label{secsol}

The solution in the above is
\begin{eqnarray}\label{anisoBHsol}
&&\mathrm{d}s^2 = -V(r)\mathrm{d}t^2 + \frac{1}{V(r)}\mathrm{d}r^2 +r^2\mathrm{d}x^2 + G(x)\Big[r^2\mathrm{d}y^2 + h(r)H(y)
\mathrm{d}z^2\Big]\,,\qquad h(r)\,, H(y) \,\text{are arbitrary}\\
&&V(r)=-\frac{\Lambda  r^2}{3}-M\,,\text{ $M>0$ for a black hole with the horizon at $r_h=\sqrt{-3M/\Lambda}$,}\\
&&G(x)=\left(b_1 e^{\sqrt{M} x}+b_2 e^{-\sqrt{M} x}\right)^2\,,\qquad \text{and $b_1\,,b_2$ and $M$ are constants.}
\end{eqnarray}
The horizon and other constant $t,r$-surfaces has the geometry described by
\begin{eqnarray}\label{anisohorizonexact}
	\mathrm{d}x^2 + \left(b_1 e^{\sqrt{M} x}+b_2 e^{-\sqrt{M} x}\right)^2 \left(\mathrm{d}y^2 +H(y)\mathrm{d}z^2\right)\,.
\end{eqnarray}
This solution exists only when $\alpha$ is fixed to the critical value
\begin{eqnarray}\label{criticalalpha2}
\alpha = -\frac{3}{4\Lambda}\,.
\end{eqnarray}
Interestingly, when $h(r)$ is set to be $r^2$, the horizon manifold can be arbitrary as shown in the paper \cite{Dotti:2007az}, including the cases of the solution (\ref{anisoBHsol}) above. 
The choice $\alpha=-3/(4\Lambda)$ is the critical point where there is one single AdS vacuum for the Einstein-Gauss-Bonnet theory\cite{Cai:2001dz}
and is the same as that in Ref.\cite{Banados:1993ur} on dimensionally continued gravity and Ref.\cite{Fan:2016zfs}, where they found the critical value of the coupling constant actually forbids kinetic fluctuations around the AdS backgrounds.
In the previous paper \cite{Peng:2021xwh} we arrived at the same choice to allow nontrival topological horizons.
The critical value of the Gauss-Bonnet coupling $\alpha$ plays a quite important role in the existence of the nontrival horizons. 

\textbf{An important question is: are the arbitrary functions $h(r)$ and $H(y)$ genuine arbitrariness, or gauge degrees of freedom which can be eliminated by coordinate transformations?}
One can look at the Ricci scalar
\begin{eqnarray}
&&R=\frac{1}{6  r^2 G(x)^2 h(r)^2 H(y)^2}\left(3 G(x) h(r)^2 \left(H'(y)^2-2 H(y) \left(2  H(y) G''(x)+H''(y)\right)\right)+3 h(r)^2 H(y)^2 G'(x)^2\right.\nonumber\\
&&\left.+ G(x)^2 H(y)^2 \left(-r^2 h'(r)^2 \left(3 M+\Lambda  r^2\right)+2 r h(r) \left(r h''(r) \left(3 M+\Lambda  r^2\right)+h'(r) \left(6 M+4 \Lambda  r^2\right)\right)+12 h(r)^2 \left(M+2 \Lambda  r^2\right)\right)\right)\,,
\nonumber\\
\label{singularRicci}
\end{eqnarray}
and it depends on $h(r)$ and $H(y)$.
\textbf{This shows that $h(r)$ and $H(y)$ are real arbitrariness which can not be eliminated by coordinate transformations.}
In general, $R$ will at least possess a $1/r^2$ divergence.
This means that there is a singularity at $r=0$.
The tensor products $R_{ab}R^{ab}$ and $R_{abcd}R^{abcd}$ both diverge at $r=0$ too.

We finally discuss the thermodynamics briefly. The temperature can be obtained by the semi-classical method of removing the conical singularity of the near-horizon geometry
\begin{eqnarray}
T=\frac{{V'(r)}}{4 \pi }\Big|_{r\to r_h} =-\frac{\Lambda  r_h}{6 \pi }\,,
\end{eqnarray}
and the entropy can be obtained by applying the Wald entropy formula \cite{Wald:1993nt,Iyer:1994ys}
\begin{eqnarray}\label{Waldentropy}
S &=& -2\pi \int_{ {horizon}} \sqrt{\hat{g}} \,\mathrm{d}^{d-1}x \frac{\partial L}{\partial R_{abcd}}{\bm \epsilon}_{ab}{\bm \epsilon}_{cd}\nonumber\\
&=&\int \mathrm{d}x  \mathrm{d}y \mathrm{d}z \,  r_h^2 \sqrt{ h(r_h)} G(x) \sqrt{H(y)}\left[+\frac{1}{4}-\frac{G''(x)}{4 M G(x)}+\frac{G'(x)^2}{16 M G(x)^2}-\frac{H''(y)}{8  M G(x) H(y)}+\frac{H'(y)^2}{16  M G(x) H(y)^2}\right]\,,
\end{eqnarray}
where the hatted $\hat{g}$ is the intrinsic metric on the horizon cross section on which the integral is defined and ${\bm \epsilon}_{ab}$ is the natural volume element on the tangent space orthogonal to the cross section. 
From Eq.(\ref{Waldentropy}) one can see that the entropy is nonzero, compared to the zero result of the ``Solv'' manifold in Ref.\cite{Peng:2021xwh}.
This shows that the same critical choice $\alpha = -{3}/{4\Lambda}$ does not always lead to vanishing entropy, which is a thought-provoking fact.

\section{The first choice with $g(r)=h(r)$: a uniqueness theorem}\label{sectheorem}
This section introduces the case when choosing $h(r) g'(r)-g(r) h'(r)=0$ in the equation (\ref{choose}).
This leads to $g(r)=h(r)$ up to a constant. For the ansatz (\ref{anisoBHansatz}), if one set $g(r)=h(r)$, then the metric becomes
\begin{eqnarray}\label{ansatzg=h}
\mathrm{d}s^2 = -V{(r)}\mathrm{d}t^2 + \frac{1}{V(r)}\mathrm{d}r^2 + f(r)\text{d}x^2+ g(r) G(x)\left( \mathrm{d}y^2 + H(y)
\mathrm{d}z^2 \right)\,.
\end{eqnarray}
Here we present a uniqueness theorem about the geometry of the possible black hole horizons.
\newtheorem{theorem}{Theorem}
\begin{theorem}
For solutions to (\ref{GBeom}) in the form (\ref{ansatzg=h}) and expected to possess anisotropic horizons, one must have $f(r)\propto g(r)$. Then the solution goes back to the cases presented in Ref.\cite{Dotti:2007az}. In that paper the horizon can have constant curvature, or possess some other types of geometry.
\end{theorem}

Proof: when $g(r)=h(r)$ the $rx$ component of (\ref{GBeom}) gives
\begin{eqnarray}\label{g=hyy}
	\frac{G'(x) \left[f(r) g'(r)-g(r) f'(r)\right] \left[(g(r)-\alpha  g'(r) V'(r)\right]}{f(r) g(r) G(x)}=0\,.
\end{eqnarray}
For this to hold, either of the two parenthesis in the numerator vanishes. 
The choice $f(r) g'(r)-g(r) f'(r)=0$ gives $g(r)\propto f(r)$. That ends the story.
If one chooses $g(r)-\alpha  g'(r) V'(r)=0$, the $rr$ component of the field equations gives
\begin{eqnarray}
\frac{1}{G(x)}(\frac{H'(y)^2}{H(y)^2}-2\frac{H''(y)}{H(y)}) \left(\frac{f'(r)}{4 f(r)  V(r) g'(r)}-\frac{1}{4 g(r)  V(r)}\right)\nonumber\\
+\frac{G'(x)^2}{G(x)^2} \left(\frac{1}{4 f(r) V(r)}-\frac{g(r) f'(r)}{4 f(r)^2 V(r) g'(r)}\right)+(\text{terms including $r$ only})=0\,.
\end{eqnarray}
For this equation to be solved, the variables $r\,,x$ and $y$ should be separated.
The only possibility is that the first and the second coefficients are in proportion:
\begin{eqnarray}
\left(\frac{f'(r)}{4 f(r)  g'(r)}-\frac{1}{4 g(r)  }\right)\propto \left(\frac{1}{4 f(r)}-\frac{g(r) f'(r)}{4 f(r)^2 g'(r)}\right)\,,
\end{eqnarray}
leading to $f(r)=g(r)$ finally.

Therefore the solution must be in the form
\begin{eqnarray}\label{nthnew}
\mathrm{d}s^2 = -V{(r)}\mathrm{d}t^2 + \frac{1}{V(r)}\mathrm{d}r^2 + f(r)\left(\text{d}x^2+ G(x)\left( \mathrm{d}y^2 + H(y)
\mathrm{d}z^2 \right)\right)\,.
\end{eqnarray}
In the paper \cite{Dotti:2007az} the authors analyzed the solution type
\begin{eqnarray}
\mathrm{d}s^2 = -V{(r)}\mathrm{d}t^2 + \frac{1}{V(r)}\mathrm{d}r^2 + f(r)\mathrm{d}\Sigma_3^2\,,
\end{eqnarray}
and it includes the equation (\ref{nthnew}). Therefore there is nothing new here for this case, and the readers can refer to that paper for further information.

\section{conclusion and discussion}

In this paper we present a set of novel $5$-dimensional black hole solutions in Einstein-Gauss-Bonnet theory with a negative cosmological constant.
The black hole's $3$-dimensional horizon cross section possess a novel anisotropic geometry: one direction loses translational invariance. 
Besides, although the spacetime is static, the geometry is not a direct product --- there is no common warping factor.
The metric of the $3$-dimensional horizon is in the form (\ref{anisohorizonexact})
and there exists $2$ arbitrary functions $h(r)$ and $H(y)$ in the metric (\ref{anisoBHsol}) i.e. degeneracies indicating some redundant degrees of freedom.
We believe that the existence of such a kind of horizon and the degeneracy are closely related to the critical value of the coupling constant $\alpha=-3/4\Lambda$ which actually forbids kinetic fluctuations around the AdS backgrounds \cite{Fan:2016zfs}.
The same choice has lead to nontrivial horizon manifolds and degeneracies in Ref.\cite{Dotti:2007az} (with a common warping factor and arbitrary horizon geometry) and \cite{Peng:2021xwh} with a solution quite similar to our new solution (\ref{anisoBHsol}) with the ``Solv'' geometry (one of the Thurston geometries \cite{Th1})
\begin{eqnarray}
	\mathrm{d}s^2 = -(-\frac{\Lambda  r^2}{3}-M)\mathrm{d}t^2 + \frac{1}{-\frac{\Lambda  r^2}{3}-M}\mathrm{d}r^2  + r^2\mathrm{d}x^2 + r^2 e^{-2\sqrt{M}x}\mathrm{d}y^2+h(r)e^{2\sqrt{M}x}\mathrm{d}z^2
	\label{solution2}
\end{eqnarray}
with $h(r)$ undetermined.
A quite important difference is that the solution (\ref{solution2}) above has $0$ Wald entropy, possibly due to the reason that the entropy is related to the quantum degrees of freedom of the black hole, if the kinetic fluctuations are forbidden, the entropy is expected to vanish \cite{Cai:2009de,Brustein:2007jj}. 
However, the solution (\ref{anisoBHsol}) we found has nonvanishing entropy (\ref{Waldentropy}), probably because for this specific background the coefficient of the kinetic terms do not vanish.

We should emphasize that although Ref.\cite{Dotti:2007az} presented a solution with totally arbitrary horizon geometry, it does not include our new solution, since our solution does not contain a common warping factor as in that paper.
Instead the geometry of the horizon is partly fixed, while there is an arbitrary function $h(r)$.

Except the solution, we also have proven that if one wants to set a common warping factor in the rest two directions for the anisotropic horizon, the solution must contain a common warping factor for all three directions on the horizon, and the solution belongs to the known cases studied in Ref.\cite{Dotti:2007az}.

The main conclusion of this paper ends here. However, these results are quite preliminary and there might exist deep principles under the surface.
The following are some important future topics we will try our best to investigate and understand.
\begin{enumerate}
	\item There may be deep connections between the \textit{critical value} of the Gauss-Bonnet coupling and the nontrivial geometry (``Solv'' geometry and anisotropy) of the horizon together with the degeneracy of the function $h(r)$ and $H(y)$. At least we have seen that in several cases they appear together.
	Our next step is to check that if similar nontrivial horizons exist in other gravity theories with higher curvature corrections at criticality such as the theories studied in Ref.\cite{Nojiri:2001aj,Lu:2011zk,Fan:2014ala}.
	It is possible that this phenomenon is quite prevalent, and the deep reason underlying the complicated differential equations deserves more investigation.
	These topics will enrich our knowledge of classification of black hole horizons as well as our understanding of higher curvature corrections of gravity theories.
	\item Whether Maxwell or other fields will affect this horizon geometry and the degeneracy is a quite interesting question.
	There are some charged examples of which the horizon possess the Thurston geometries \cite{Arias:2017yqj,Bravo-Gaete:2017nkp,Faedo:2019rgo}.
	Inspired by these examples, one may simply add charge into our solutions and check whether these solutions still exist and what changes will take place.
    During this work, an interesting paper \cite{Yang:2023nnk} constructing nontrival horizons with scalar fields in $4$ dimensions appear. This shows that for other types of matter there would also be interesting topics. 
	Meanwhile, when the matter exists, one can also try to explore the stability analysis like that in Ref.\cite{Cai:2013cja,Cao:2021sty}.
	\item From the point of view of AdS/CFT, our solution (\ref{anisoBHsol}) implies a CFT living on a boundary with one special direction --- the translational invariance and anisotropy are broken, on the level of the spacetime background. 
	Interestingly, this boundary is only in the spatial directions, not the temporal direction.
	The physical meaning and possible applications of this boundary is an interesting topic.
\end{enumerate}

\section{Acknowledgements}
I should especially give my thanks to Prof. Rong-Gen Cai. Without him this work couldn't be done.
I also would like to thank Prof. Li-Ming Cao, Li Li, Jin-Bo Yang, Yi-Jie Zhang and Hyat Huang for helpful discussions.
This work was supported by the National Natural Science Foundation of China with Grant No. 11947029 and No.12265001 and the East China University of Technology Research Foundation for Advanced Talents (No. DHBK2019198).


\begin{thebibliography}{99}
	


\bibitem{Hawking:1971vc}
S.~W.~Hawking,
Commun. Math. Phys. \textbf{25}, 152-166 (1972)
doi:10.1007/BF01877517
		

		
		
\bibitem{Myers:1986un}
R.~C.~Myers and M.~J.~Perry,
Annals Phys. \textbf{172}, 304 (1986)
doi:10.1016/0003-4916(86)90186-7
		

\bibitem{Emparan:2001wn}
R.~Emparan and H.~S.~Reall,
Phys. Rev. Lett. \textbf{88}, 101101 (2002)
doi:10.1103/PhysRevLett.88.101101
[arXiv:hep-th/0110260 [hep-th]].

\bibitem{Galloway:2005mf}
G.~J.~Galloway and R.~Schoen,
Commun. Math. Phys. \textbf{266}, 571-576 (2006)
doi:10.1007/s00220-006-0019-z
[arXiv:gr-qc/0509107 [gr-qc]].

\bibitem{Emparan:2008eg}
R.~Emparan and H.~S.~Reall,
Living Rev. Rel. \textbf{11}, 6 (2008)
doi:10.12942/lrr-2008-6
[arXiv:0801.3471 [hep-th]].

\bibitem{Obers:2008pj}
N.~A.~Obers,
Lect. Notes Phys. \textbf{769}, 211-258 (2009)
doi:10.1007/978-3-540-88460-6\_6
[arXiv:0802.0519 [hep-th]].

\bibitem{Lemos:1994xp}
J.~P.~S.~Lemos,
Phys. Lett. B \textbf{353}, 46-51 (1995)
doi:10.1016/0370-2693(95)00533-Q
[arXiv:gr-qc/9404041 [gr-qc]].

\bibitem{Lemos:1995cm}
J.~P.~S.~Lemos and V.~T.~Zanchin,
Phys. Rev. D \textbf{54}, 3840-3853 (1996)
doi:10.1103/PhysRevD.54.3840
[arXiv:hep-th/9511188 [hep-th]].

\bibitem{Birmingham:1998nr}
D.~Birmingham,
Class. Quant. Grav. \textbf{16}, 1197-1205 (1999)
doi:10.1088/0264-9381/16/4/009
[arXiv:hep-th/9808032 [hep-th]].

\bibitem{Aminneborg:1996iz}
S.~Aminneborg, I.~Bengtsson, S.~Holst and P.~Peldan,
Class. Quant. Grav. \textbf{13}, 2707-2714 (1996)
doi:10.1088/0264-9381/13/10/010
[arXiv:gr-qc/9604005 [gr-qc]].


\bibitem{Cai:1996eg}
R.~G.~Cai and Y.~Z.~Zhang,
Phys. Rev. D \textbf{54}, 4891-4898 (1996)
doi:10.1103/PhysRevD.54.4891
[arXiv:gr-qc/9609065 [gr-qc]].









	
		
\bibitem{Cadeau:2000tj}
C.~Cadeau and E.~Woolgar,
Class. Quant. Grav. \textbf{18}, 527-542 (2001).

\bibitem{Th1} W.P.\ Thurston, {\it Three-Dimensional Geometry and 
	Topology}, ed.\ S.\ Levy (Princeton University Press, Princeton, 1997).









\bibitem{Lovelock:1971yv} 
D.~Lovelock,
J.\ Math.\ Phys.\  {\bf 12}, 498 (1971).
doi:10.1063/1.1665613


\bibitem{Gross:1986iv} 
D.~J.~Gross and E.~Witten,
Nucl.\ Phys.\ B {\bf 277}, 1 (1986).
doi:10.1016/0550-3213(86)90429-3

\bibitem{Zumino:1985dp} 
B.~Zumino,
Phys.\ Rept.\  {\bf 137}, 109 (1986).
doi:10.1016/0370-1573(86)90076-1

\bibitem{Cai:2001dz} 
R.~G.~Cai,
Phys.\ Rev.\ D {\bf 65}, 084014 (2002)
doi:10.1103/PhysRevD.65.084014
[hep-th/0109133].


\bibitem{Cvetic:2001bk}
M.~Cvetic, S.~Nojiri and S.~D.~Odintsov,
Nucl. Phys. B \textbf{628} (2002), 295-330
doi:10.1016/S0550-3213(02)00075-5
[arXiv:hep-th/0112045 [hep-th]].


\bibitem{Peng:2021xwh}
Y.~Peng,
Phys. Rev. D \textbf{104}, no.8, 084004 (2021)
doi:10.1103/PhysRevD.104.084004
[arXiv:2105.08482 [gr-qc]].

\bibitem{Fan:2016zfs}
Z.~Y.~Fan, B.~Chen and H.~Lu,
Eur. Phys. J. C \textbf{76}, no.10, 542 (2016)
doi:10.1140/epjc/s10052-016-4389-x
[arXiv:1606.02728 [hep-th]].


\bibitem{Arenas-Henriquez:2019rph}
G.~Arenas-Henriquez, R.~B.~Mann, O.~Miskovic and R.~Olea,
Phys. Rev. D \textbf{100}, no.6, 064038 (2019)
doi:10.1103/PhysRevD.100.064038
[arXiv:1905.10840 [hep-th]].


\bibitem{Vegh:2013sk}
D.~Vegh,
[arXiv:1301.0537 [hep-th]].

\bibitem{Davison:2013jba}
R.~A.~Davison,
Phys. Rev. D \textbf{88}, 086003 (2013)
doi:10.1103/PhysRevD.88.086003
[arXiv:1306.5792 [hep-th]].



\bibitem{Blake:2013bqa}
M.~Blake and D.~Tong,
Phys. Rev. D \textbf{88}, no.10, 106004 (2013)
doi:10.1103/PhysRevD.88.106004
[arXiv:1308.4970 [hep-th]].

\bibitem{Blake:2013owa}
M.~Blake, D.~Tong and D.~Vegh,
Phys. Rev. Lett. \textbf{112}, no.7, 071602 (2014)
doi:10.1103/PhysRevLett.112.071602
[arXiv:1310.3832 [hep-th]].







\bibitem{Andrade:2013gsa}
T.~Andrade and B.~Withers,
JHEP \textbf{05}, 101 (2014)
doi:10.1007/JHEP05(2014)101
[arXiv:1311.5157 [hep-th]].


\bibitem{Cai:2014znn}
R.~G.~Cai, Y.~P.~Hu, Q.~Y.~Pan and Y.~L.~Zhang,
Phys. Rev. D \textbf{91}, no.2, 024032 (2015)
doi:10.1103/PhysRevD.91.024032
[arXiv:1409.2369 [hep-th]].

\bibitem{Davison:2014lua}
R.~A.~Davison and B.~Gout\'eraux,
JHEP \textbf{01}, 039 (2015)
doi:10.1007/JHEP01(2015)039
[arXiv:1411.1062 [hep-th]].



\bibitem{Xu:2015rfa}
J.~Xu, L.~M.~Cao and Y.~P.~Hu,
Phys. Rev. D \textbf{91}, no.12, 124033 (2015)
doi:10.1103/PhysRevD.91.124033
[arXiv:1506.03578 [gr-qc]].

\bibitem{Cao:2015cza}
L.~M.~Cao and Y.~Peng,
Phys. Rev. D \textbf{92}, no.12, 124052 (2015)
doi:10.1103/PhysRevD.92.124052
[arXiv:1509.08738 [hep-th]].

\bibitem{Cao:2015cti}
L.~M.~Cao, Y.~Peng and Y.~L.~Zhang,
Phys. Rev. D \textbf{93}, 124015 (2016)
doi:10.1103/PhysRevD.93.124015
[arXiv:1511.04967 [hep-th]].
\bibitem{Chen:2019zlg}
F.~Chen, S.~F.~Wu and Y.~Peng,
JHEP \textbf{07}, 072 (2019)
doi:10.1007/JHEP07(2019)072
[arXiv:1903.02672 [hep-th]].


\bibitem{Sabin:2014bua}
C.~Sabin, D.~E.~Bruschi, M.~Ahmadi and I.~Fuentes,
New J. Phys. \textbf{16}, 085003 (2014)
doi:10.1088/1367-2630/16/8/085003
[arXiv:1402.7009 [quant-ph]].

\bibitem{Biasi:2019eap}
A.~Biasi, J.~Mas and A.~Serantes,
JHEP \textbf{05}, 161 (2019)
doi:10.1007/JHEP05(2019)161
[arXiv:1903.05618 [hep-th]].


\bibitem{An:2019opz}
Y.~S.~An, R.~G.~Cai, L.~Li and Y.~Peng,
Phys. Rev. D \textbf{101}, no.4, 046006 (2020)
doi:10.1103/PhysRevD.101.046006
[arXiv:1909.12172 [hep-th]].


\bibitem{Dotti:2007az}
G.~Dotti, J.~Oliva and R.~Troncoso,
Phys. Rev. D \textbf{76}, 064038 (2007)
doi:10.1103/PhysRevD.76.064038
[arXiv:0706.1830 [hep-th]].




\bibitem{Banados:1993ur}
M.~Banados, C.~Teitelboim and J.~Zanelli,
Phys. Rev. D \textbf{49}, 975-986 (1994)
doi:10.1103/PhysRevD.49.975
[arXiv:gr-qc/9307033 [gr-qc]].


\bibitem{Wald:1993nt} 
R.~M.~Wald,
Phys.\ Rev.\ D {\bf 48}, no. 8, R3427 (1993).

\bibitem{Iyer:1994ys} 
V.~Iyer and R.~M.~Wald,
Phys.\ Rev.\ D {\bf 50}, 846 (1994).



\bibitem{Cai:2009de}
R.~G.~Cai, L.~M.~Cao and N.~Ohta,
Phys. Rev. D \textbf{81}, 024018 (2010)
doi:10.1103/PhysRevD.81.024018
[arXiv:0911.0245 [hep-th]].


\bibitem{Brustein:2007jj}
R.~Brustein, D.~Gorbonos and M.~Hadad,
Phys. Rev. D \textbf{79}, 044025 (2009)
doi:10.1103/PhysRevD.79.044025
[arXiv:0712.3206 [hep-th]].








\bibitem{Nojiri:2001aj}
S.~Nojiri and S.~D.~Odintsov,
Phys. Lett. B \textbf{521} (2001), 87-95
[erratum: Phys. Lett. B \textbf{542} (2002), 301]
doi:10.1016/S0370-2693(01)01186-8
[arXiv:hep-th/0109122 [hep-th]].



\bibitem{Lu:2011zk}
H.~Lu and C.~N.~Pope,
Phys. Rev. Lett. \textbf{106}, 181302 (2011)
doi:10.1103/PhysRevLett.106.181302
[arXiv:1101.1971 [hep-th]].


\bibitem{Fan:2014ala}
Z.~Y.~Fan and H.~Lu,
Phys. Rev. D \textbf{91}, no.6, 064009 (2015)
doi:10.1103/PhysRevD.91.064009
[arXiv:1501.00006 [hep-th]].



\bibitem{Arias:2017yqj}
R.~E.~Arias and I.~Salazar Landea,
JHEP \textbf{12}, 087 (2017)
doi:10.1007/JHEP12(2017)087
[arXiv:1708.04335 [hep-th]].




\bibitem{Bravo-Gaete:2017nkp}
M.~Bravo-Gaete and M.~Hassaine,
Phys. Rev. D \textbf{97}, no.2, 024020 (2018)
doi:10.1103/PhysRevD.97.024020
[arXiv:1710.02720 [hep-th]].






\bibitem{Faedo:2019rgo}
F.~Faedo, D.~A.~Farotti and S.~Klemm,
JHEP \textbf{12}, 151 (2019)
doi:10.1007/JHEP12(2019)151
[arXiv:1908.07421 [hep-th]].


\bibitem{Yang:2023nnk}
J.~Yang,
[arXiv:2301.01709 [gr-qc]].

\bibitem{Cai:2013cja}
R.~G.~Cai and L.~M.~Cao,
Phys. Rev. D \textbf{88}, 084047 (2013)
doi:10.1103/PhysRevD.88.084047
[arXiv:1306.4927 [gr-qc]].



\bibitem{Cao:2021sty}
L.~M.~Cao and L.~B.~Wu,
Phys. Rev. D \textbf{103}, no.6, 064054 (2021)
doi:10.1103/PhysRevD.103.064054
[arXiv:2101.02461 [gr-qc]].






		
%
%

		
		

		
		




		






		
	\end{thebibliography}
\end{document}